\documentclass[twocolumn, aps, superscriptaddress]{revtex4-1}

\usepackage{amssymb, amsmath}
\usepackage{amsmath}
\usepackage{graphicx}
\usepackage{siunitx}
\usepackage{hyperref}
\usepackage{color}
\hypersetup{colorlinks=false, citecolor=blue, urlcolor=blue, linkcolor=black}

\begin{document}

\title{Low-Temperature, Dry Transfer-Printing of a Patterned Graphene Monolayer}%

\author{Sugkyun Cha}%
\thanks{These authors contributed equally.}
\affiliation{Program in Nano Science and Technology, Graduate School of Convergence Science and Technology, Seoul National University, Seoul 151-742, Republic of Korea}

\author{Minjeong Cha}%
\thanks{These authors contributed equally.}
\affiliation{Program in Nano Science and Technology, Graduate School of Convergence Science and Technology, Seoul National University, Seoul 151-742, Republic of Korea}

\author{Seojun Lee}%
\affiliation{Program in Nano Science and Technology, Graduate School of Convergence Science and Technology, Seoul National University, Seoul 151-742, Republic of Korea}

\author{Jin Hyoun Kang}%
\affiliation{Department of Chemistry, Seoul National University, Seoul 151-747, Republic of Korea}

\author{Changsoon Kim}%
\email{changsoon@snu.ac.kr}
\affiliation{Program in Nano Science and Technology, Graduate School of Convergence Science and Technology, Seoul National University, Seoul 151-742, Republic of Korea}
\affiliation{Advanced Institutes of Convergence Technology, Suwon, Gyeonggi 443-270, Republic of Korea}

\date{\today}

\begin{abstract}
Graphene has recently attracted much interest as a material for flexible, 
transparent electrodes or active layers in electronic and photonic devices. 
However, realization of such graphene-based devices is limited due to 
difficulties in obtaining patterned graphene monolayers on top of materials
that are degraded when exposed to a high-temperature or wet process.
We demonstrate a low-temperature, dry process capable of transfer-printing a patterned graphene monolayer grown on Cu foil onto a target substrate using an elastomeric stamp. 
A challenge in realizing this is to obtain a high-quality graphene layer on a 
hydrophobic stamp made of poly(dimethylsiloxane), which is overcome by 
introducing two crucial modifications to the conventional wet-transfer method -- the use 
of a support layer composed of Au and the decrease in surface tension of the liquid bath. 
Using this technique, patterns of a graphene monolayer were transfer-printed on 
poly(3,4-ethylenedioxythiophene):polystyrene sulfonate and MoO$_3$, 
both of which are easily degraded when exposed to an aqueous or aggressive patterning
process. We discuss the range of application of this technique, which is currently limited
by oligomer contaminants, and possible means to expand it by eliminating the
contamination problem.
\end{abstract}

\maketitle

\section{Introduction}
Graphene, a one-atom-thick layer of carbon atoms arranged in a hexagonal lattice, has outstanding electrical~\cite{Novoselov2004, CastroNeto2009} and mechanical~\cite{Liu2007, Lee2008} properties,
as well as high optical transmittance~\cite{Nair2008}. For this reason, many electronic and 
photonic devices employing graphene, as either an active layer or a transparent electrode,
have been demonstrated, such as light-emitting diodes (LEDs)~\cite{Meyer2014, Kim2011}, 
solar cells~\cite{Wang2011, Park2012}, field-effect transistors~\cite{Di2008},
photodetectors~\cite{Baierl2011}, touch screens~\cite{Bae2010}, terahertz wave
modulators~\cite{Sensale-Rodriguez2012, Gao2014, Li2015}, and Schottky junction
devices~\cite{Yang2012, Tongay2012}. In many such demonstrations, a graphene layer
has been deposited by transferring it onto a device substrate following the
conventional wet-transfer method, where a graphene--polymer bilayer floating on a water
bath is scooped by the substrate~\cite{Li2009}. And when the patterning of graphene
layers is required, it has mostly been performed after graphene transfer, typically
using photolithography followed by reactive-ion etch (RIE)~\cite{Lee2011, Shi2013}. 
However, this method of obtaining patterned graphene layers -- the wet-transfer and
subsequent patterning process -- has only a limited range of applications, where
graphene layers must be deposited and patterned, when necessary, prior to deposition of 
any material that is too fragile to withstand a wet, high-temperature, or plasma process.
Notable, practically important examples of such materials are organic semiconductors~\cite{Forrest2004} and organometal trihalide perovskite compounds~\cite{Snaith2013}

Attention, therefore, has been focused on development of dry-transfer techniques~\cite{Kang2012}.
For example, a graphene layer grown on a Cu layer on a donor substrate can be directly
transferred onto a target substrate, by delaminating the graphene--Cu interface
when the target substrate in contact with the graphene layer is peeled off from the donor substrate~\cite{Yoon2012}. However, for selective delamination, the target substrate
needs to be coated with an epoxy adhesion layer, which makes this technique
unsuitable for high-performance electronic devices: for example, it cannot be applied
to fabrication of an LED with a top graphene electrode, since the adhesion layer
in this case would be placed in the device interior, just beneath the graphene electrode,
impeding efficient charge injection. Another approach is to transfer-print a graphene
layer coated with a `self-release' layer from an elastomeric stamp onto a target 
substrate~\cite{Song2013}, where reliable transfer is achieved by choosing
an appropriate self-release layer that assures the selective delamination at the interface
between that and the elastomer. Although the transfer process itself is dry, 
removing the self-release layer transferred along with the graphene is typically
achieved with an organic solvent, ultimately limiting applications of this method.
Jung \textit{et al.} demonstrated a technique capable of transferring graphene
monolayers without an adhesion or a self-release layer~\cite{Jung2014}.
In this mechano-electro-thermal process, complete transfer, instead, requires
application of high temperature ($\ge$ \SI{160}{\degreeCelsius}) and voltage ($\ge$ \SI{600}{V})
while a graphene layer grown on Cu foil is pressed onto a target substrate. 
\begin{figure*}[t]
\includegraphics[width=1.8\columnwidth]{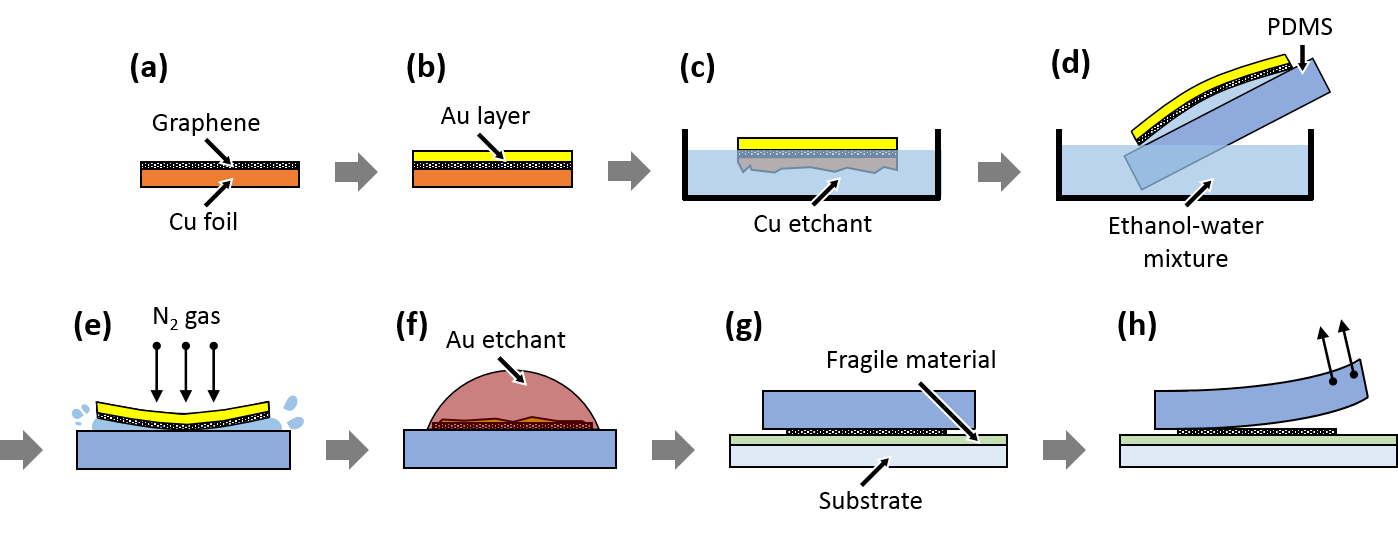}
\caption{\textbf{Transfer-printing of a CVD-grown graphene monolayer.}
The process begins with a graphene monolayer grown on Cu foil (a).
After depositing a support layer composed of Au (b), the Cu layer is etched by floating the sample on a Cu etchant bath (c).
After transferring the resulting graphene--Au bilayer to a bath comprising a mixture of water and ethanol, the bilayer was scooped up with a PDMS stamp (d).
Next, the sample is dried using a N$_2$ gun (e), followed by heat treatment on a hot plate.
Removing the Au layer (f) completes the fabrication of the stamp coated with the graphene monolayer, which is then gently pressed onto a target substrate coated with a material that can be easily damaged by a wet process (g).
The graphene monolayer is transferred onto the substrate as the stamp is peeled off from the substrate (h).
Between (g) and (h), the stamp and substrate are kept in conformal contact for \SI{1}{\hour}, followed by heat treatment on a hot plate.}
\label{fig:process_schematic}
\end{figure*}

Here, we demonstrate a low-temperature, dry transfer process capable of transfer-printing a patterned
graphene monolayer onto a target substrate that can be damaged or degraded
by a wet, plasma or high-temperature process. In this process, a graphene monolayer
on Cu foil, which is grown by chemical vapor deposition (CVD) and then patterned
using a conventional lithographic process, is transferred onto a stamp made of poly(dimethylsiloxane) (PDMS), and
subsequently transfer-printed from the stamp onto the target substrate.
The graphene transfer from Cu foil to PDMS is achieved using the conventional
wet-transfer process~\cite{Li2009}, with the following two modifications:
the use of Au, instead of poly(methyl methacrylate) (PMMA), as a material for the support
layer, and the decrease in surface tension of the liquid bath using a water-ethanol mixture.
These modifications are critical in preventing defect formation in a graphene monolayer
during its transfer onto a PDMS stamp, thereby leading to a minimum sheet resistance of
\SI{573}{\ohm\per sq} for a graphene monolayer transfer-printed onto a glass substrate.
Furthermore, we demonstrate transfer-printing of patterned graphene monolayers on
poly(3,4-ethylenedioxythiophene):polystyrene sulfonate (PEDOT:PSS) and MoO$_3$,
which are representative examples of organic electronic materials and practically important metal oxides~\cite{Girotto2011}, respectively, that are usually damaged or
degraded when exposed to aqueous or aggressive patterning processes.
The morphological and elemental characterizations of the surfaces of transfer-printed
graphene show the existence of contaminants that are likely to be siloxane oligomers
transferred from the PDMS stamp. We discuss the current range of application of this
technique and possible means to expand it by eliminating the contamination problem.

\section{Transfer-Printing of a Patterned Graphene Layer}

To transfer a graphene monolayer onto a target substrate that can be damaged or
degraded by a wet or high-temperature process (Fig.~\ref{fig:process_schematic}),
we first transfer a CVD-grown graphene onto a PDMS stamp following the conventional
wet-transfer method (a to f): by scooping up, with the PDMS stamp, a graphene--support bilayer floating on liquid. After the support layer is removed by chemical
etching, the graphene is transfer-printed on a target substrate (g to h). The first part of this
process (a to f), although seemingly similar to the conventional wet-transfer technique~\cite{Li2009}, has two distinct features, which are crucial to obtain a high-quality
graphene monolayer on a target substrate.



First, as a support layer material, we use thermally deposited Au, instead of PMMA,
which is mostly widely used for this purpose in the wet-transfer method~\cite{Li2009a}.
PDMS, the material chosen for a stamp owing to its mechanical and chemical properties 
suitable for various transfer-printing techniques~\cite{Xia1998}, swells when immersed in
an organic solvent~\cite{Lee2003} that can dissolve the PMMA support layer, such as
acetone and chloroform. When this occurs, the graphene monolayer cracks, creating a large number of defects (Supplementary Fig.~ S1). On the contrary, the use of a Au support layer allows one to obtain a high-quality
graphene monolayer on PDMS, since Au can be removed using an aqueous etchant,
which does not swell PDMS. 

Second, for the liquid on which the graphene--Au bilayer floats and from which
it is scooped with a PDMS stamp [Fig.~\ref{fig:process_schematic}(d)], 
we use an ethanol--water mixture, instead of water commonly used in the
conventional wet-transfer technique. This is to decrease the surface tension of the liquid.
In the conventional case, after the graphene--support bilayer is scooped with a hydrophilic substrate [as in Fig.~\ref{fig:process_schematic}(d)], a thin layer of water is present
throughout the graphene--substrate interface, providing sufficient lubrication
at that interface. As a result, when the sample is blow-dried using a N$_2$ gun,
the graphene and substrate form a conformal contact without wrinkles throughout
the substrate, as the water is laterally displaced [Supplementary Fig.~S2(a)].
Since the surface of a PDMS stamp is hydrophobic, which is favorable for
reliable transfer of a graphene monolayer onto a target substrate via stamping
(g to h in Fig.~\ref{fig:process_schematic}), the use of water bath in Fig.~\ref{fig:process_schematic}(d) leads to a discontinuous lubrication layer between
the bilayer and substrate, as schematically shown in Supplementary Fig.~S2(b).
Therefore, blow-drying in this case results into bursting of trapped water droplets,
tearing the graphene monolayer. This can be effectively prevented by using an
ethanol--water mixture as the liquid bath, which sufficiently wets the PDMS surface to
provide a continuous lubrication layer [Supplementary Fig.~S2(c)].

When patterning of graphene is required, a conventional patterning process, such as O$_2$ RIE of graphene using photoresist patterned by photolithography as an etch mask~\cite{Lee2011, Shi2013}, is performed before Step (b) in Fig.~\ref{fig:process_schematic}.
Then, performing the remaining processes [Step (b) to (h)], one can obtain a patterned
graphene monolayer on a target substrate. This pre-transfer patterning of graphene
allows one to avoid possible damage to the fragile material that is likely to occur,
when a process such as photolithography~\cite{Lee2011, Shi2013}, RIE~\cite{Lee2011, Shi2013}, or laser ablation~\cite{Kalita2011} is performed after the graphene is transferred to the target substrate.

\section{Results and Discussion}

\begin{figure*}[t]
\includegraphics[width=1.8\columnwidth]{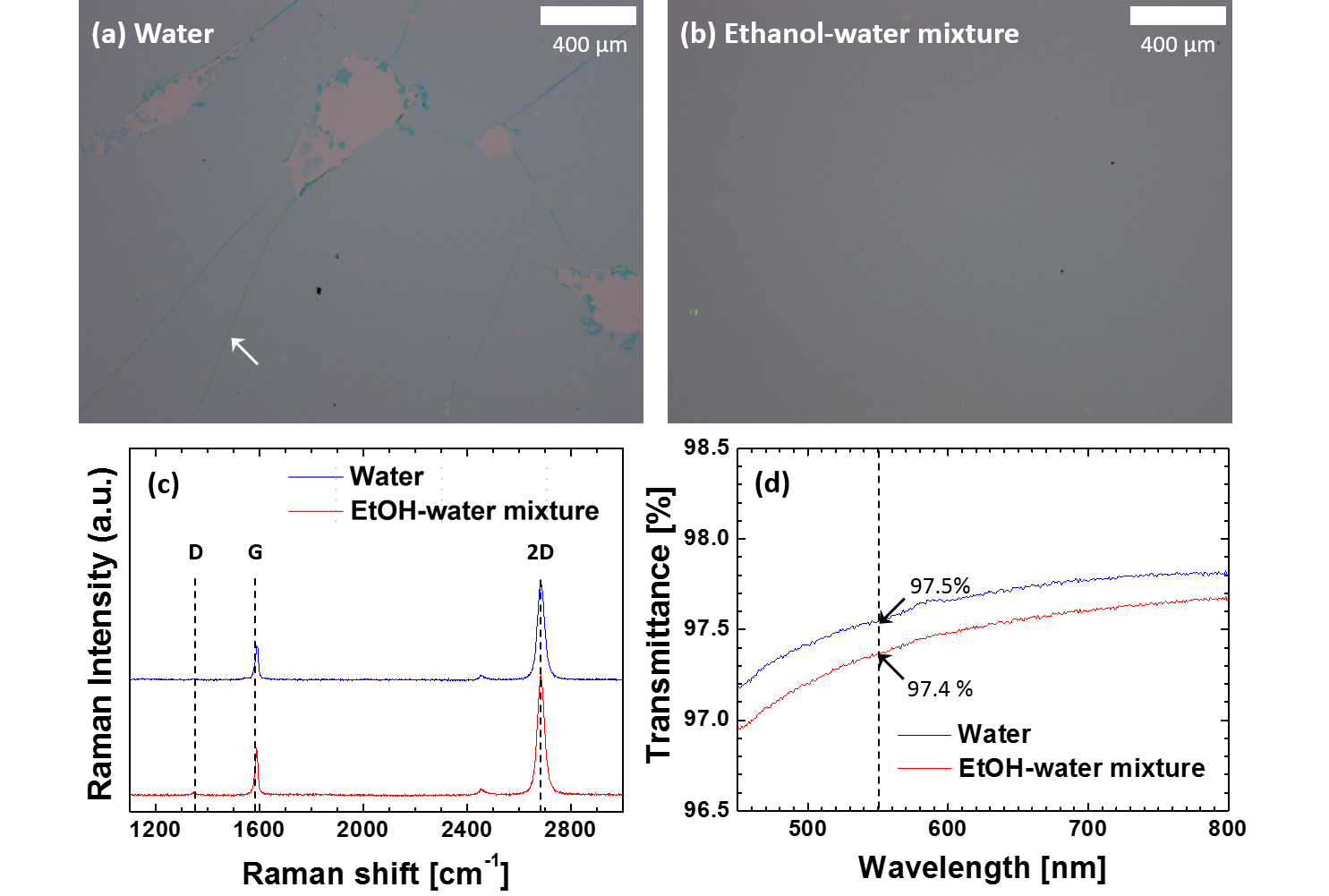}
\caption{\textbf{Effect of surface tension on the quality of transfer-printed graphene monolayers.} (a,b) Optical microscope images of graphene monolayers transfer-printed on Si substrates pre-coated with SiO$_2$. The bath used in step (d) in Fig.~\ref{fig:process_schematic} is composed of water in (a) and a water--ethanol mixture in (b). Along a line indicated by the arrow in (a), the graphene monolayer is folded, which arises from a corresponding wrinkle in the graphene--Au bilayer formed due to insufficient wetting of the PDMS surface by the bilayer. (c) Raman spectra of the graphene monolayers on the Si/SiO$_2$ substrates. (d) Optical transmittance spectra of the graphene monolayers transferred on glass substrates. Each curve was obtained by averaging over five samples.}
\label{fig:water_vs_mixture}
\end{figure*}

To show that the surface tension of a liquid used in Step (d) in Fig.~\ref{fig:process_schematic} is a critical factor determining the quality of
transfer-printed graphene, we transfer-printed a graphene monolayer on a Si
substrate coated with a 285-nm-thick SiO$_2$ layer following a process described
in Fig.~\ref{fig:process_schematic}, while varying the liquid bath: in one set of
experiments, we used water, and in the other, a water--ethanol mixture
(\SI{30}{\percent} water and \SI{70}{\percent} ethanol by volume). When water bath was used, although
the entire graphene sheet (\SI{1.3}{\centi\meter} by \SI{1.3}{\centi\meter}) was seemingly well-transferred,
a closer observation revealed that there are randomly distributed irregular-shaped
holes where graphene is absent, as shown in Fig.~\ref{fig:water_vs_mixture}(a).
The density of these defects is approximately \SI{10}{\per\centi\meter\squared}. When the PDMS stamp was observed by an optical microscope after Step (f) in Fig.~\ref{fig:process_schematic}, it was found that similar defects, albeit smaller in size,
were present (Supplementary Fig.~S3), indicating that the defects are formed while transferring the graphene layer onto the PDMS surface and are exacerbated during the transfer-printing onto the substrate.
As described in the previous section, the defects arise from insufficient wetting of
the PDMS surface by water. Since PDMS is hydrophobic, immediately after
a graphene--Au bilayer is scooped by a PDMS stamp, water dewets the PDMS surface
in several locations, making the bilayer form contacts to the PDMS surface that is only
locally conformal [Supplementary Figs.~S2(b) and S4(a)]. As the sample is blow-dried
using a N$_2$ gun, these locally conformal contacts laterally expand, generating
narrow wrinkles with water droplets trapped inside, as shown in the right image of
Supplementary Fig.~S4(a). We speculate that further application of N$_2$ pressure
causes the water droplets to burst, resulting into defects such as that shown in
Supplementary Fig.~S3. In fact, as shown in Fig.~\ref{fig:water_vs_mixture}(a),
the locations of many defects in the graphene transferred onto the substrate
seem to coincide with the intersection of the wrinkles, where relatively large water
droplets are expected to form: the linear regions in Fig.~\ref{fig:water_vs_mixture}(a)
indicated by the white arrow are where the graphene monolayer is folded, which
results from the wrinkles in the graphene--Au bilayer. In contrast, when the
ethanol--water mixture was used, its lower surface tension ($\simeq$ \SI{25}{dyn\per\centi\meter} at \SI{23}{\degreeCelsius}~\cite{Hansen1998} allows a continuous lubrication layer to form between
the graphene and PDMS surfaces, providing effective ``decoupling'' of the bilayer
from the PDMS surface. Therefore, no wrinkles, except a few with much smaller
heights, were observed in the graphene--Au bilayer on the PDMS stamp [Supplementary
Fig.~S4(b)]. We found that mild baking at \SI{40}{\degreeCelsius} removes these wrinkles,
resulting into the flat graphene--Au bilayer that is globally conformal to the PDMS stamp,
and consequently, successful transfer-printing of the graphene monolayer was
achieved without defects, as shown in Fig.~\ref{fig:water_vs_mixture}(b).

The sheet resistance ($R_\text{sh}$) was measured for graphene monolayers
transfer-printed on glass substrates, using the van der Pauw method~\cite{VanderPauw1958}. The size of the graphene monolayers are approximately
\SI{1.3}{\centi\meter} by \SI{1.3}{\centi\meter}. In the following, a graphene monolayer transfer-printed onto
a final substrate from a PDMS stamp onto which a graphene--Au bilayer was
scooped from a bath of water and the ethanol--water mixture are referred to as
$G_\text{H$_2$O}$ and $G_\text{EtOH--H$_2$O}$, respectively.
For $G_\text{H$_2$O}$, the sheet resistance, averaged over five samples
($\langle{R}_\text{sh}\rangle$) is \SI{3119}{\ohm\per sq}, with a minimum equal to
\SI{2664}{\ohm\per sq}.
In contrast, for $G_\text{EtOH--H$_2$O}$,  $\langle{R}_\text{sh}\rangle$ is \SI{914}{\ohm\per sq}, with a minimum being \SI{573}{\ohm\per sq}. Figure~\ref{fig:water_vs_mixture}(c) shows
Raman spectra of graphene monolayers shown in Figs.~\ref{fig:water_vs_mixture}(a)
and (b), where for $G_\text{H$_2$O}$ they
were obtained from defect-free regions. The spectra show that, for both cases,
(i) each Raman peak occurs at the same location (D: \SI{1344}{\per\centi\meter}, 2D: \SI{2686}{\per\centi\meter}, G: \SI{1588}{\per\centi\meter}), (ii) the height of the D peaks is negligible, and
(iii) the 2D/G peak ratios are larger than 2.7, confirming that
the transfer-printed graphene is indeed a monolayer~\cite{Ferrari2006}. This result indicates that
significantly larger values of $R_\text{sh}$ for $G_\text{H$_2$O}$, in comparison to that for $G_\text{EtOH--H$_2$O}$, are due not to the properties of graphene in defect-free regions, but to large-scale defects as shown in Fig.~\ref{fig:water_vs_mixture}(a),
which has been prevented by decreasing the surface tension
in the case of $G_\text{EtOH--H$_2$O}$.

Figure~\ref{fig:water_vs_mixture}(d) shows the optical transmission spectra of
$G_\text{H$_2$O}$ and $G_\text{EtOH--H$_2$O}$ transfer-printed
on a 0.7-mm-thick glass substrate, averaged over five samples for each case.
Transmittance ($T$), plotted on the $y$-axis, is the intensity of the optical beam
transmitted through a glass/graphene sample normalized to that transmitted
through a glass substrate. The size of the optical beam at the sample location
was approximately \SI{2}{\milli\meter} by \SI{8}{\milli\meter}.
For both $G_\text{H$_2$O}$ and $G_\text{EtOH--H$_2$O}$, the values of $T$
are consistent with what was previously measured for a graphene monolayer
on a quartz substrate~\cite{Bae2010}.
The value of $T$ for $G_\text{H$_2$O}$ is slightly higher than
that for $G_\text{EtOH--H$_2$O}$, primarily because the absence of graphene
in the defects in $G_\text{H$_2$O}$ allow more light to be transmitted.
Under this hypothesis, the ratio of total area of the defects to the entire area
of the graphene sheet ($\alpha$), can be estimated as
$\alpha = (T_\text{H$_2$O} - T_\text{EtOH--H$_2$O})/(1-T_\text{EtOH--H$_2$O})$,
where  $T_\text{H$_2$O}$ and  $T_\text{EtOH--H$_2$O}$ are transmittance of
$G_\text{H$_2$O}$ and $G_\text{EtOH--H$_2$O}$, respectively.
The value of  $\alpha$ calculated at each wavelength in Fig.~\ref{fig:water_vs_mixture}(d)
ranges from \SI{6}{\percent} to \SI{8}{\percent}, which is consistent with our estimation based on optical
microscope images.

As expected, successful transfer-printing of graphene requires a defect-free
graphene monolayer that is globally conformal to a PDMS stamp.
Our proposed technique achieves this with the water--ethanol mixture,
which provides a continuous lubrication layer, and with an Au support layer,
which allows for its removal without swelling PDMS. Alternatively, one may
attempt to obtain a defect-free graphene monolayer on a PDMS stamp
by pressing the stamp onto a graphene layer grown on Cu foil and then etching away
the Cu foil by floating the Cu/graphene/PDMS on a bath of a Cu etchant.
Since the surface of Cu foil commonly used in CVD growth of graphene
typically has corrugations on the micron scale~\cite{Li2009a}, the PDMS attached
to the graphene in this case is in contact with the graphene only partially.
As a result, subsequent processes such as N$_2$ blow-dry and transfer-printing
tend to cause defects in the graphene layer, as shown in Supplementary Fig.~S5.
In fact, it was previously reported that $R_\text{sh}$ of a transfer-printed graphene
monolayer by this approach was \SI{4}{\kilo\ohm\per sq}, even with a self-release layer
inserted for reliable graphene transfer~\cite{Song2013}.

\begin{figure*}[t]
\includegraphics[width=1.8\columnwidth]{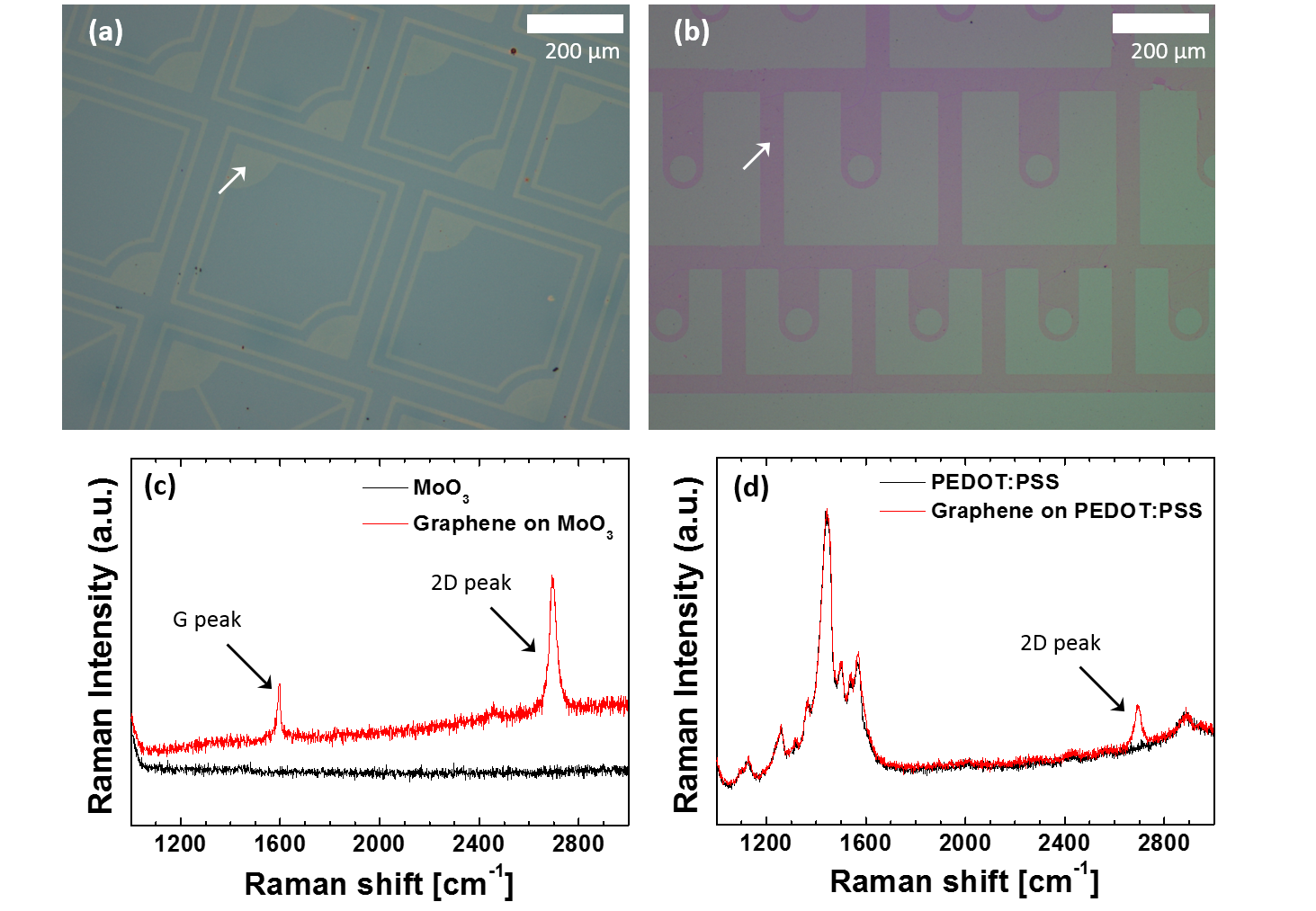}
\caption{\textbf{Characterizations of graphene monolayer patterns transfer-printed on materials that can be damaged by a wet process.} (a,b) Optical microscope images of graphene monolayer patterns transfer-printed on (a) MoO$_3$ and (b) PEDOT:PSS. The arrows in (a) and (b) indicate regions covered by graphene monolayers. (c, d) Raman spectra (red) of the patterned graphene on (c) MoO$_3$ and (d) PEDOT:PSS. Also shown are the Raman spectra of the substrates without graphene (black).
}
\label{fig:patterning}
\end{figure*}

To fabricate practical electronic devices where graphene is used as active layers
or electrodes, the patterning of graphene is required. 
Our technique, described in Fig.~\ref{fig:process_schematic}, can achieve this
with a simple modification: the process begins with a patterned graphene on Cu foil
in Step (a), instead of an unpatterned graphene layer. In our current demonstration,
we first prepared a patterned graphene monolayer on Cu foil by etching
unpatterned graphene grown on Cu foil by O$_2$ RIE using a photoresist etch mask
patterned by photolithography. Next, the patterned graphene was transfer-printed on
a Si/SiO$_2$ substrate coated with MoO$_3$ or PEDOT:PSS, 
both of which are susceptible to degradation when exposed to an aqueous condition
or aggressive patterning process. Figures~\ref{fig:patterning}(a) and (b) are optical
micrographs of the substrates, where patterned graphene monolayers were
transfer-printed in regions indicated by the arrows,
showing that the patterns defined on photomasks were replicated
in the transfer-printed graphene monolayers.
The widths of the smallest features -- lines in Fig.~\ref{fig:patterning}(a) and arcs in Fig.~\ref{fig:patterning}(b) -- are \SI{10}{$\mu$\meter} and \SI{15}{$\mu$\meter}, respectively,
which are identical, within the resolution of the optical imaging system used
($\sim$ \!\SI{0.5}{$\mu$\meter}), to those of the corresponding features on the photomask.
A closer observation of the pattern edge using a field emission scanning electron microscope (FE-SEM)
 revealed that it is not straight on the nanoscale,
with an ``edge resolution'' of \SI{50}{\nano\meter}, which is probably attributed to the edge resolution
of the photomask patterns and/or limitation of the photolithography process
(Supplementary Fig.~S7).
From this, together with the fact that previously demonstrated transfer-printing-based
patterning techniques can create patterns whose size is well below
100~nm~\cite{Kim2002}, we expect that our technique is capable of creating
sub-micrometer graphene patterns, if a nanopatterning process,
for example electron-beam~\cite{Withers2011}, nanoimprint~\cite{Liang2010},
or nanosphere~\cite{Liu2011} lithography, is employed, instead of photolithography.
Raman spectra obtained from the graphene transfer-printed on the MoO$_3$
show the distinct G and 2D peaks, with the 2D/G intensity ratio of 2.5,
and the negligible D peak, suggesting that the quality of the graphene is comparable to
that in Fig.~\ref{fig:water_vs_mixture}(b).
For the case of the graphene transfer-printed onto the PEDOT:PSS,
the peaks associated with graphene, except the 2D peak, cannot be identified
due to the overlap with Raman spectra of PEDOT:PSS.

\begin{figure*}[t]
\includegraphics[width=1.6\columnwidth]{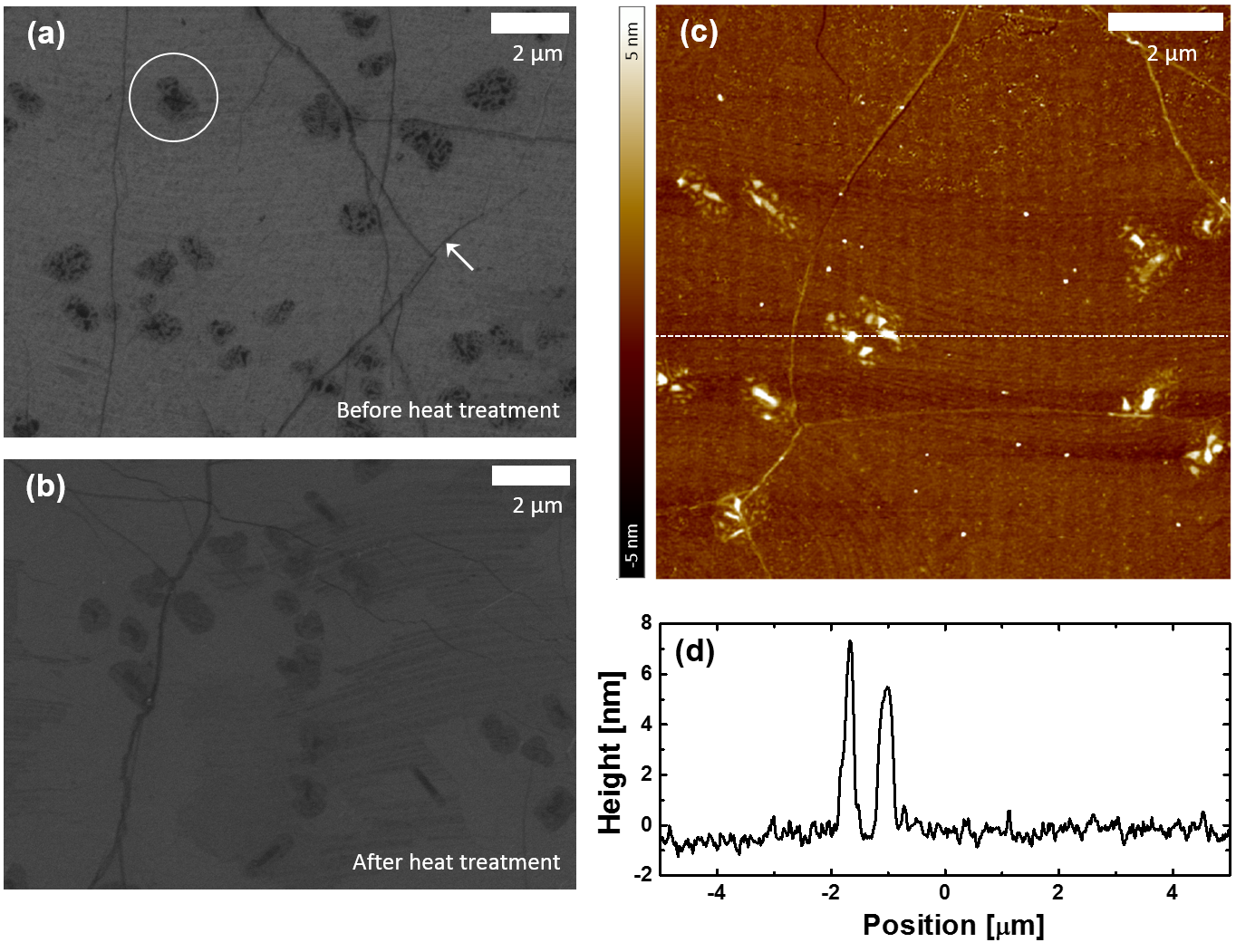}
\caption{\textbf{Morphological characterizations of transfer-printed graphene monolayers.} (a, b) Scanning electron microscope images of the graphene surface before (a) and after (b) annealing at 400~$^\circ$C under H$_2$ and Ar. (c) Atomic force microscope image of the sample used in (a). (d) Height profile of the surface measured along the white dotted line in (c).}
\label{fig:sem_afm}
\end{figure*}

Next, we observed the surface of the transfer-printed graphene on
a Si/SiO$_2$ substrate, using a FE-SEM and an atomic force microscope (AFM).
As shown in Fig.~\ref{fig:sem_afm}(a), irregularly shaped dark patches,
as enclosed by a white circle, are randomly distributed throughout the surface.
Also shown are the dark lines, as marked by the white arrow.
These two features, patches and lines, are commonly found in the transferred graphene CVD-grown on Cu foil -- with the former and latter attributed to graphene multilayers and wrinkles, respectively~\cite{Li2009a} -- and hence are not caused by our transfer technique.
It is also shown that in the patches, there are darker spots with diameters
of approximately \SI{150}{\nano\meter}.
The surface profile measured using an AFM along the white dotted line in Fig.~\ref{fig:sem_afm}(c) shows that the spots have heights as high as
approximately \SI{7}{\nano\meter} [Fig.~\ref{fig:sem_afm}(d)].
To identify the origin of the dark spots, elemental analysis was carried out
using a scanning transmission electron microscope capable of energy dispersive
x-ray spectroscopy (STEM-EDS).
In order to prepare a sample for this analysis, a graphene layer was
transfer-printed from a PDMS stamp onto a Si/SiO$_2$ substrate coated
with a PEDOT:PSS separation layer, and then transferred onto a lacey carbon
TEM grid using the conventional wet-transfer method (see the Method Section for the experimental detail).
An EDS spectrum obtained from a region shown in Supplementary Fig.~S8(a) shows that, in addition to carbon, silicon atoms are present on the graphene surface [Supplementary Fig.~S8(b)].
Given many previous reports showing that uncured siloxane oligomers were present on PDMS surfaces~\cite{Glasmastar2003, Song2013}, it is highly likely
that the dark spots on the graphene surface are siloxane residues
that have been transferred from the PDMS stamp.
This speculation was further supported by the fact that the dark spots can be eliminated
by annealing the sample at \SI{400}{\degreeCelsius} under H$_2$ and Ar, as shown in Fig.~\ref{fig:sem_afm}(b)~\cite{Allen2009}.
The AFM measurements [Fig.~\ref{fig:sem_afm}(c) and (d)] show that the surface
in the background, that is, regions away from the patches and lines,
is much rougher than that of a clean graphene surface~\cite{Pirkle2011},
suggesting that the oligomer residues are also present throughout the surface,
not only on the multilayer regions.

The morphological and elemental characterizations of the surface of transfer-printed
graphene discussed above help determine the range of application of our technique
in its current form.
Since the oligomer residues are likely to be present only on the top surface, that is,
the graphene surface that used to be in contact with the PDMS, 
our technique can be applied to fabrication of (i) devices where only the bottom surface of the graphene electrode is involved in injection
or collection of charge carriers, such as LEDs and solar cells, made of organic semiconductors~\cite{Beck2015} or organometal trihalide perovskite compounds~\cite{You2015}, with top graphene electrodes,
and (ii) devices whose graphene electrodes are used to establish electric fields
without charge carrier transport, such as thin-film transistors
with graphene gate electrodes~\cite{Park2011} and
terahertz wave modulators~\cite{Sensale-Rodriguez2012, Gao2014, Li2015}.
Meanwhile, when charge carrier injection or collection occurs
in both sides of the graphene layer, such as in tandem LEDs and solar cells
where it is part of the interlayers, our technique is not applicable.
Therefore, expanding the range of application of our technique
by eliminating the oligomer contamination, possibly with the following modification,
is important future work: replacing PDMS with other stamp material that can be
completely cured; or depositing a blocking layer on the PDMS surface to prevent possible transfer of uncured oligomers onto the graphene surface, with a potential candidate being a pressure sensitive adhesive layer~\cite{Choi2015, Kim2015}.

\section{Conclusion}

In summary, we have developed a low temperature, dry process capable of
transfer-printing a patterned graphene monolayer grown on Cu foil on a target substrate. Two features distinct from the conventional wet-transfer method~\cite{Li2009} 
-- the use of a support layer composed of Au, instead of PMMA, and the decrease in surface tension of the liquid bath on which a graphene--Au bilayer floats --
allow one to obtain a graphene monolayer on a PDMS stamp without defects that would otherwise arise.
Subsequently, the graphene is transfer-printed from the stamp onto a target substrate.
The characteristics of a graphene monolayer transfer-printed using our technique are comparable to those obtained with the conventional wet-transfer method,
with a sheet resistance as low as \SI{573}{\ohm \per sq} and optical transmittance of
\SI{97.4}{\percent} at \SI{550}{\nano\meter}.
In addition, with pre-transfer patterning of graphene on Cu foil using conventional
patterning processes, our technique is capable of creating graphene monolayer
patterns on materials that are easily degraded when exposed to high-temperature processes, organic solvents, or aqueous chemicals.
As an example, using photolithography followed by reactive-ion etch to pattern graphene monolayers on Cu foil
and then transfer-printing them, we have obtained graphene monolayer patterns
on MoO$_3$ and PEDOT:PSS, with the smallest feature size and edge resolution of
$\simeq$ \SI{10}{$\mu$\meter} and \SI{50}{\nano\meter}, respectively.
Immediate application areas of this technique include organic electronic devices
whose top electrodes are composed of graphene.
Moreover, by eliminating siloxane oligomer residues on graphene
using alternate stamp material, the technique can be further applied to devices whose graphene electrodes are in their interiors, such as tandem LEDs and solar cells.
Finally, with possible appropriate modification, it may also be applied to dry-transfer of other two-dimensional materials, including boron nitride~\cite{Zhang2015} and molybdenum disulfide~\cite{Shi2012}.

\section*{Methods}

\subsection*{CVD-grown graphene monolayer}
A graphene monolayer on Cu foil was grown in a CVD system consisting of a tubular quartz reactor and a furnace.
Experimental details described in Ref.~\cite{Bae2010} were closely followed except the following: Cu foil was annealed under a 5 sccm flow of H$_2$ at \SI{20}{\milli Torr}, and during growth, the reactor was filled with a mixture of CH$_4$ and H$_2$ at a total pressure of \SI{150}{\milli Torr}, whose flow rates are 35 and \SI{5}{sccm}, respectively.

\subsection*{Low-temperature, dry transfer-printing process}
Low-temperature, dry transfer of graphene monolayers was carried out by following processes described in Fig.~\ref{fig:process_schematic}.
To form a support layer on a graphene monolayer on Cu foil, a 200-nm-thick Au layer was deposited by thermal evaporation in high vacuum (\SI[per-mode=symbol]{1}{\angstrom\per \second}, $\sim$\! \SI{d-7}{Torr}) (b).
To etch away the Cu foil, the Cu/graphene/Au multilayer was floated on an ammonium persulfate solution, prepared by dissolving \SI{10}{\gram} of ammonium persulfate (Sigma Aldrich) in \SI{500}{\milli\liter} of water (c).
After the etch was completed, the graphene--Au bilayer was scooped with a glass slide, and then transferred on a bath of water to remove residual ammonium persulfate.
Next, the graphene--Au bilayer was moved onto a bath composed of an ethanol--water mixture (70 vol \% ethanol and 30 vol \% water), from which the bilayer was scooped by and transferred onto a PDMS stamp (d).
The sample was then blow dried using a N$_2$ gun (e), and was further dried on a hot plate at \SI{40}{\degreeCelsius} for more than \SI{4}{\hour}.
The Au support layer was etched using an ammonium iodide solution (LAE-202, Cowon Innotech. Inc.) (f), after which the PDMS/graphene sample was rinsed with water.
After water droplets on the sample were blown away using a N$_2$ gun, the graphene-coated PDMS stamp was gently pressed onto a target substrate, inducing intimate contact throughout the substrate area (g).
Before separation of the stamp from the substrate, the sample was stored at room temperature for \SI{1}{\hour} under a pressure of \SI{9,9}{\kilo\pascal}, and then placed on a hot plate at \SI{70}{\degreeCelsius} for \SI{10}{\minute} without application of pressure.
Finally, the stamp was carefully peeled off from the substrate (h), resulting in the transfer-printed graphene monolayer on the target substrate.

\subsection*{Transfer-printing of patterned graphene layers}
In this process, a graphene monolayer on Cu foil was first patterned using conventional photolithography and reactive-ion etch, as described in Supplementary Fig.~S6.
A 1.5-$\mu$m-thick photoresist (AZ GXR-601, \SI{14}{cP}) was spin-coated on a Cu/graphene sample, and then patterned by photolithography.
The patterned graphene on Cu foil was obtained, when the graphene in the areas not covered by the photoresist was etched by reactive ion etch in O$_2$ (\SI{100}{W}, \SI{0.1}{Torr}, \SI{20}{\second}, \SI{50}{sccm}).
Performing the processes described in Fig.~\ref{fig:process_schematic} with this sample, rather than unpatterned graphene on Cu foil, we transfer-printed a patterned graphene monolayer on a target substrate coated with a 75-nm-thick PEDOT:PSS or a 20-nm MoO$_3$ layer.
The target substrate was a 500-$\mu$m-thick Si substrate pre-coated with a 285-nm-thick thermal SiO$_2$ layer, and the PEDOT:PSS (Heraeus) and MoO$_3$ (LTS Chemical Inc.) layers were deposited by spin-coating (\SI{3000}{rpm, \SI{30}{\second}}) and thermal evaporation in high vacuum (\SI[per-mode=symbol]{1}{\angstrom\per\second}, $\sim$\! \SI{d-7}{Torr}), respectively.

\subsection*{Sample preparation for the elemental analysis}
Samples for the elemental analysis were prepared following the processes described in Supplementary Fig.~S9.
After a graphene monolayer was transfer-printed from a PDMS stamp onto a PEPOT:PSS layer using our transfer method (a), a layer of PMMA was deposited on the graphene layer by spin coating at \SI{3000}{rpm} for \SI{30}{\second} (b).
The PMMA solution was prepared by dissolving PMMA (\SI{138}{\milli\gram}, Sigma Aldrich) into chlorobenzene (\SI{3}{\milli\liter}, Sigma Aldrich).
The sample was then immersed into a water bath, separating the PMMA--graphene bilayer from the Si/SiO$_2$ substrate as the PEDOT:PSS layer was dissolved (c).
Next, the bilayer was transferred to another water bath and kept floating on it for more than \SI{24}{\hour} to ensure that PEDOT:PSS remaining on the graphene surface was removed.
Then, the bilayer was scooped with a lacey carbon TEM grid (Ted Pella, Inc.)
(d), after which the grid was placed on a hot plate at \SI{40}{\degreeCelsius} for more than \SI{2}{\hour}.
Finally, the PMMA layer was removed by acetone (e), resulting in the graphene monolayer on the TEM grid (f).

\subsection*{Characterization of transfer-printed graphene layers}
The surface morphology of transfer-printed graphene layers was characterized using a FE-SEM (JSM-6700F, JEOL) and an AFM (Dimension Edge, Bruker).
The sheet resistance was measured using a source meter (2400, Keithley) and a multimeter (34410A, Agilent).
The Raman spectra were obtained using a confocal Raman microscope (inVia, Renishaw) with an excitation wavelength of \SI{514.5}{\nano\meter} emitted from an Ar laser.
An ultraviolet--visible spectrophotometer (Lambda 35, Perkin Elmer) was used to measure the optical transmittance spectra.
The elemental analysis was carried out using a STEM equipped with an EDS (JEM-2100F, JEOL).

\section*{Acknowledgements}
This work was supported by the Global Frontier R\&D Program on the Center for Multiscale Energy System (Grant No. 2011-0031561) and the Center for THz-Bio Application Systems (Grant No. 2009-0083512), both by the National Research Foundation under the Ministry of Science, ICT, and Future Planning, Korea.

\section*{Author contribution statement}
M.\,C. and C.\,K. conceived the main idea of the transfer-printing process, and designed the experiments; S.\,C. performed CVD-growth of graphene used in all experiments, prepared and characterized graphene patterns on MoO$_3$ and PEDOT:PSS, and performed AFM and SEM measurements, and elemental characterization;
S.\,C. and M.\,C. carried out the remaining experiments, except the acquisition of the Raman spectra, which was performed by J.\,H.\,K.; S.\,C., M.\,C., and C.\,K. analyzed the data; C.\,K, S.\,C. and S.\,L. wrote the manuscript.

\section*{Competing financial interests}
The authors declare no competing financial interests.

\bibliographystyle{apsrev4-1}
%

\widetext
\clearpage
\begin{center}
\textbf{\large Supplementary Material for \\[5mm ]Low-Temperature, Dry Transfer-Printing of a Patterned Graphene Monolayer\\[1.5cm]}
\end{center}

\setcounter{equation}{0}
\setcounter{figure}{0}
\setcounter{table}{0}
\setcounter{page}{1}
\makeatletter
\renewcommand{\theequation}{S\arabic{equation}}
\renewcommand{\thefigure}{S\arabic{figure}}
\renewcommand{\bibnumfmt}[1]{[S#1]}
\renewcommand{\citenumfont}[1]{S#1}

\begin{figure}[h]
\includegraphics[width=0.5\columnwidth]{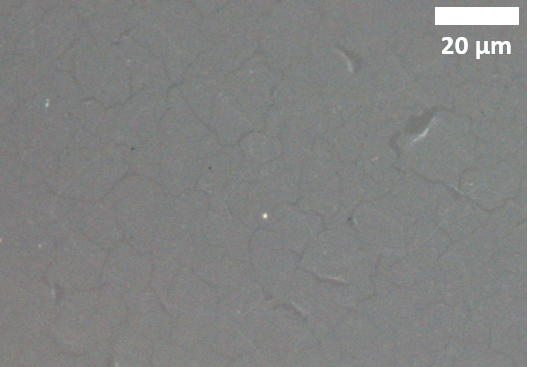}
\caption{Optical microscope image of a graphene monolayer transferred onto a PDMS stamp via the conventional wet-transfer method that uses a PMMA support layer.}
\label{fig:graphene_on_PDMS}
\end{figure}

\begin{figure*}[b]
\includegraphics[width=0.95\columnwidth]{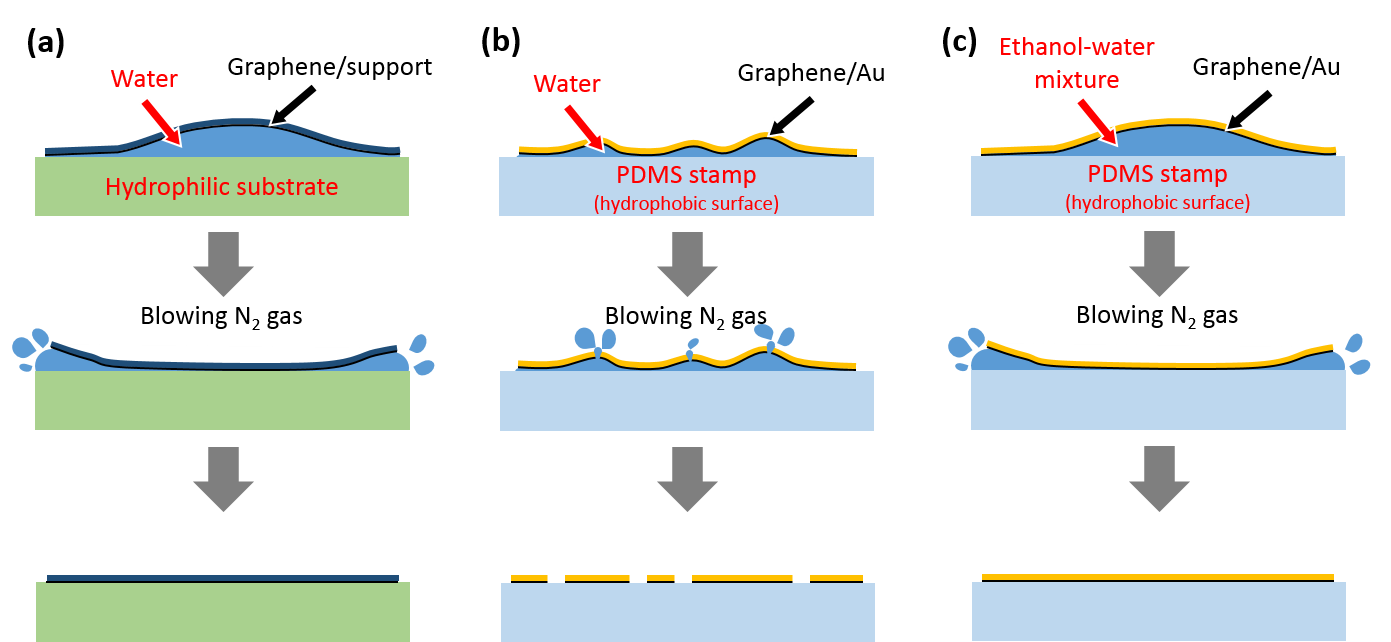}
\caption{\textbf{Importance of wetting of a substrate in determining the quality of a graphene--support bilayer on the substrate: schematic illustration.} (a) In the conventional wet-transfer case, sufficient wetting of a hydrophilic substrate by water leads to conformal contact between the graphene and the substrate without wrinkles when the sample is blow-dried using N$_2$ gas. (b) When a PDMS stamp, whose surface is hydrophobic, is used instead of a hydrophilic substrate, water does not form a continuous layer between the graphene and the PDMS stamp. Consequently, a blow-drying process in this case causes water droplets trapped between the graphene and the PDMS to burst, damaging the graphene--Au bilayer. (c) In our method, the surface tension of the bath is decreased using a mixture of water and ethanol, which sufficiently wets the PDMS surface. As a result, the graphene--Au bilayer whose quality is comparable to that in the conventional wet-transfer is obtained on the PDMS stamp. }
\label{fig:wetting_cartoon}
\end{figure*}

\begin{figure}[h]
\includegraphics[width=0.5\columnwidth]{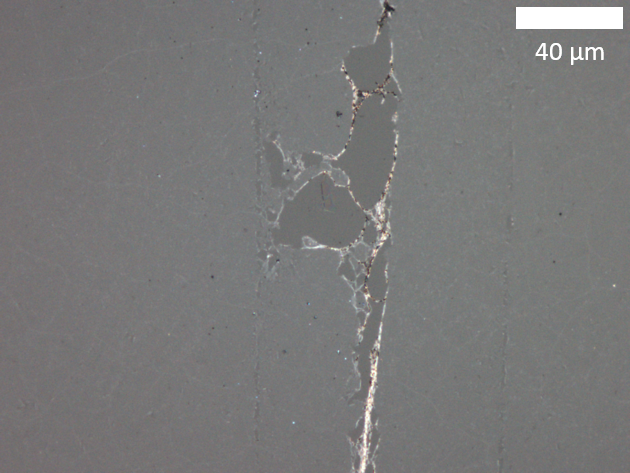}
\caption{Optical microscope image of a graphene layer transferred on a PDMS stamp using a water bath as in Step (d) in Fig. 1. }
\label{fig:graphene_on_PDMS_water}
\end{figure}

\begin{figure}[b]
\includegraphics[width=0.95\columnwidth]{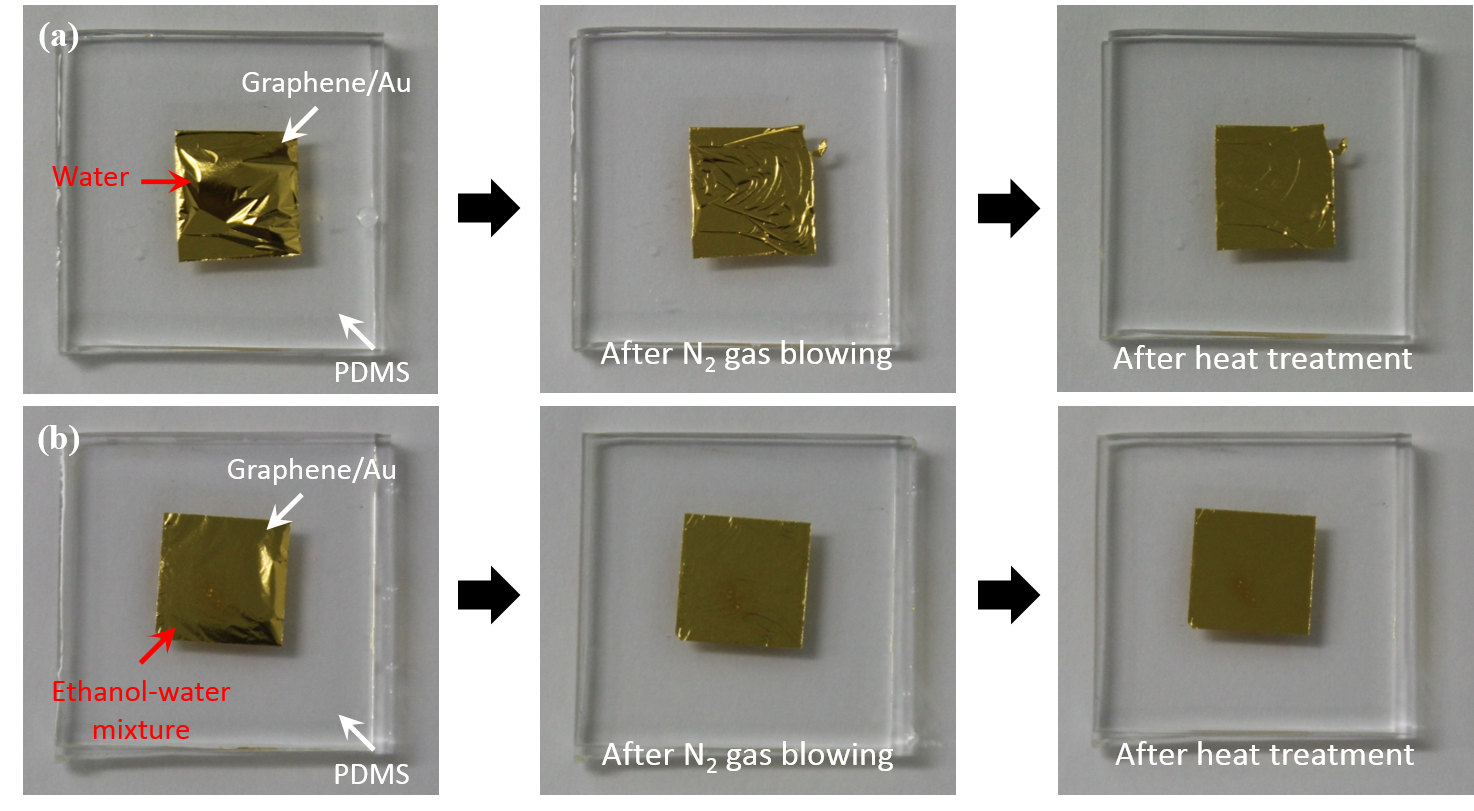}
\caption{\textbf{Importance of wetting of a substrate in determining the quality of a graphene--Au bilayer on the substrate: digital images.} These are digital images of a graphene--Au bilayer scooped up from a bath with a PDMS stamp, as shown in Fig. 1(d). (a) When a water bath is used, water dewets the PDMS surface in several locations before a blow-dry process (left). After blow-drying the sample with a N$_2$ gun, the bilayer has many wrinkles, in which water is trapped (center). Although mild annealing on a hot plate at \SI{40}{\degreeCelsius} for \SI{4}{\hour} decreases the heights and number of the wrinkles as it removes the residual water, the wrinkles cannot be completely eliminated (right). (b) In contrast, when the bilayer was scooped up from a mixture of water and ethanol, the mixture liquid forms a continuous lubrication layer between the bilayer and the PDMS stamp throughout the surface (left). As a result, N$_2$ flow aiming at the center of the bilayer displaces the liquid outward, resulting in conformal contact between the bilayer and the PDMS, almost throughout the surface (center). Small number of wrinkles with smaller heights than those in (a) can be removed after heat treatment on a hot plate at \SI{40}{\degreeCelsius} for \SI{4}{\hour} (right). }
\label{fig:wetting_photos}
\end{figure}

\begin{figure}[t]
\includegraphics[width=0.8\columnwidth]{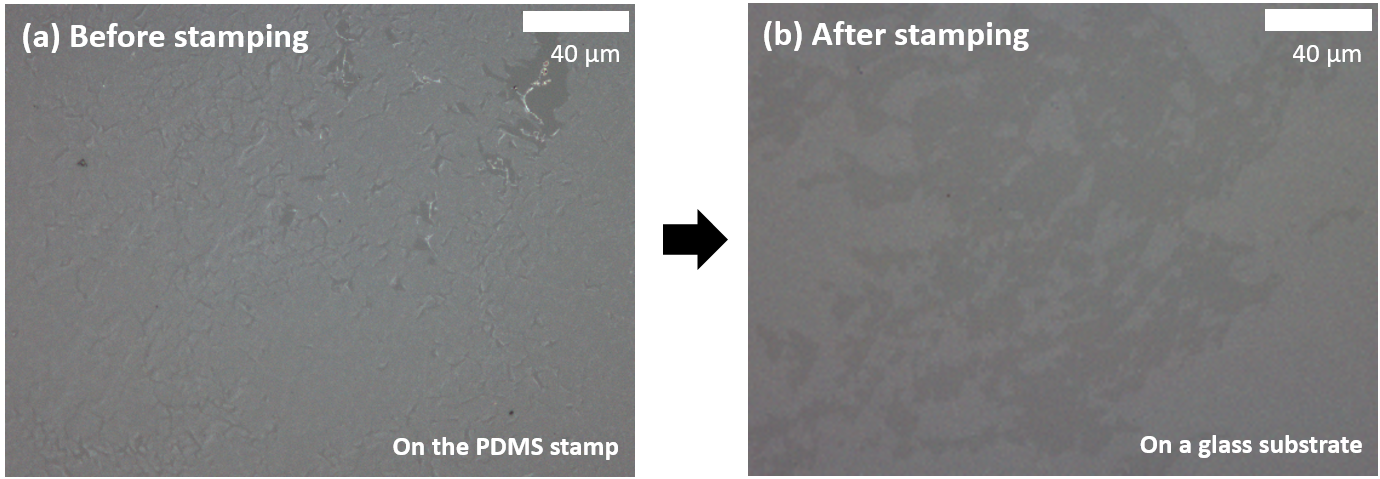}
\caption{(a) Optical microscope image of a PDMS stamp coated with a graphene layer, showing defects in the graphene. Here, the graphene layer was deposited on the PDMS by pressing the stamp onto the graphene layer on Cu foil, and floating the resulting Cu/graphene/PDMS structure on a Cu etchant solution to etch the Cu layer. (b) Optical microscope image of the graphene layer transfer-printed on a glass substrate from a stamp such as that shown in (a), showing a large number of defects.}
\label{fig:PDMS_attach}
\end{figure}

\begin{figure}[ht]
\includegraphics[width=0.8\columnwidth]{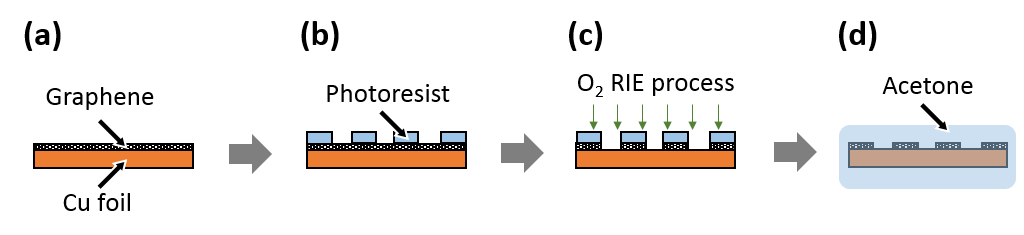}
\caption{\textbf{Pre-transfer patterning of graphene layers.} (a) CVD-grown graphene layer on Cu foil. (b) A photoresist layer on the Cu/graphene patterned by conventional photolithography. (c) Reactive-ion etch of the graphene. (d) Removal of the photoresist by acetone.}
\label{fig:rie}
\end{figure}

\begin{figure}[b]
\includegraphics[width=0.5\columnwidth]{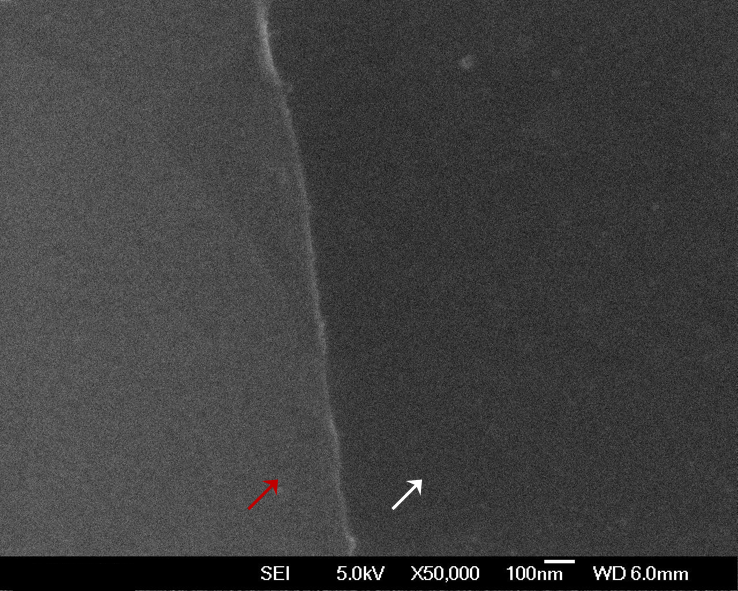}
\caption{SEM image of a graphene layer transfer-printed on MoO$_3$, showing a pattern edge resolution of approximately 50~nm. The regions indicated by red and white arrows are the graphene and MoO$_3$ surfaces, respectively. }
\label{fig:edge}
\end{figure}

\begin{figure}[t]
\includegraphics[width=0.9\columnwidth]{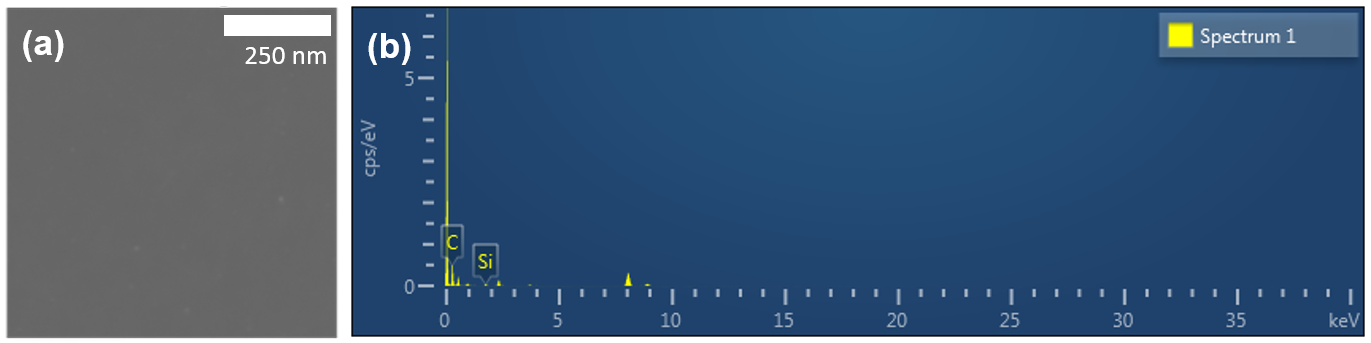}
\caption{\textbf{Elemental characterization of a transfer-printed graphene layer.} (a) Scanning transmission electron microscope image of the region where an energy dispersive x-ray spectroscopy (EDS) was performed. (b) EDS spectrum obtained from the region shown in (a).
}
\label{fig:EDX}
\end{figure}

\begin{figure}[h]
\includegraphics[width=0.9\columnwidth]{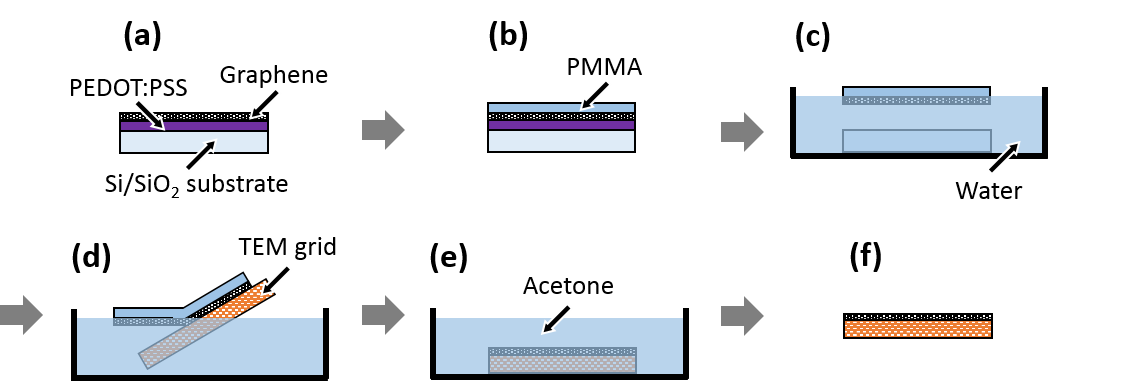}
\caption{\textbf{Sample preparation for elemental analysis.} (a) Transfer-printing of a graphene layer using the technique proposed in this paper onto a Si/SiO$_2$/PEDOT:PSS. (b) Deposition of a PMMA layer by spin coating. (c) Lifting off the graphene--PMMA bilayer by dissolving the PEDOT:PSS layer in water. (d) Transferring the graphene--PMMA bilayer onto a lacey carbon grid. (e) Removal of the PMMA layer using acetone. (f) The graphene layer transfer-printed onto the target substrate is now placed on the lacey carbon grid for the elemental analysis.}
\label{fig:EDX_sample_prep}
\end{figure}

\end{document}